# ***Risk and Profitability Analysis of FNJN Holdings Inc.***

Jack Colleran, John Sun, Benny Pickle, and Harris Kim

December 4, 2018

Finjan Holdings, Inc. (Finjan) is a cyber security firm who holds patents for softwares that help detect potential threats online. Started in 1995, the firm believes in innovation, investments and strategic partnerships as a business model[2].

Finjan has three separate businesses, Finjan Mobile, Finjan Blue and CybeRisk. These companies target different consumers and specialize in different softwares. According to Finjan's 2018 10-K, "we generate revenues and related cash flows by granting intellectual property licenses for the use of patented technologies that we own" and suing companies who infringe on their patents. They make and enforce licensing agreements through negotiation and litigation. Licenses sold are paid with a lump sum license fee instead of royalties which are not necessarily paid immediately. When negotiating these fees, Finjan takes into account how much the firm has made off their software already and how much they will make in the future.

In 2017, Finjan completed four separate lawsuits. According to footnotes in their 2017 10-K, these lawsuits earned Finjan approximately $90 million dollars. Of this, $65 million is to be paid in 2018 and will not show up on the 2017 financial statements. Finjan earns a significant portion of its revenue from lawsuits and license agreements in which they lack customers. The customers that they do have tend to be cyber security firms who have licensed their software. However, Finjan mentions in their 10-K that would like to expand their CybeRisk business which has a consumer customer base. This poses as a risk as it is a change to their current business model. This could result in potential cost increases in the future as the firm is not used to this kind of business with a stable consumer base.

COGS only represented 12% of their revenues in 2017, which is typical for cybersecurity companies. Net income including SGA expenses total to 55% of revenue in 2017.

[2] Finjan, 2018 10-k, 3

However, such large margins are not normally the case. In 2016, Finjan had $18 million in revenue but only $350 thousand in net income. Even though there was a huge difference in net income year over year, gross margins were nearly constant. Over the last three years have all been between 80% and 88%. A lucrative year in litigation and licencing fees is the difference between a successful and unsuccessful year for the firm. The company's assets are mostly intangible in the form of patents. Thus, their revenue is derived from lawsuits and licencing fees.

This business model makes the footnotes of the 10-K extremely important. Without knowing where the firm stands in their legal proceedings, it is very hard to paint an accurate picture of the firm. Looking at Finjan's financial statements does little to project their future performance. In order to judge future success, the continued supply of footnotes are needed. For example, the footnotes showing lawsuits on the horizon that can provide a windfall for Finjan, is far more useful than looking at previous years net income, for example. Finjan's inventory turnover ratio reveals nothing.

A potentially issue arising for Finjan is that they are dependent on expiring patents and have not acquired enough new ones according to page 13 of the 10-K. The business will lose its monopoly on certain technologies unless they change their business model to a non-patent trolling enterprise or acquire new patents. The usefulness of these technologies, and the names of barriers to entry of manufacturing such technologies even if anyone with capital may produce them remains a subject for further study. A competition to hire scientists to develop new technologies, to offer financial incentives may arise during the patent expiration period. Even non-expiring patents based on older technologies have value converging to zero because new technologies will replace them[3]. In perspective, they have 3 patent-pending. For future 10-Ks, a

-

table of their patents along with their patent families, countries may or may not be released depending on its news perception value to FNJN.

Their 10-K continues to mention how they are heavily reliant on "lengthy and uncertain" legal process. If the lawsuits do not go their way, Finjan is highly susceptible to incur a losses.

Finjan recognises the future of their company could be in jeopardy. This is the reason they are investing in CybeRisk. They are doing this through cash investments while avoiding raising debt. They believe that raising debt would limit their flexibility if they were to incur a loss during a long period of time.

[3] Finjan, 2018 10 k, 13

**Resources and Funding Sources**

Finjan follows policies of keeping track of possible impairment of long-term assets. Finjan creates an allowance for doubtful accounts. They estimate the amount of the allowance for doubtful accounts using predictors including "historical experience, credit quality, the age of the accounts receivable balances, and current economic conditions that may affect a customer's ability to pay."[4] However, Finjan did not record bad debt expense for 2017, 2016, and 2015 as it was nonexistent, indicating that they follow up on getting its receivables.

When accounting for accumulated depreciation of property and equipment, it is calculated on the straight-line method over the estimated useful lives of the related assets, which range from 3 to 7 years. Costs of betterments and additions are accounted with expenditures being expense in the period until the items of property and equipment are sold or retired. When sold/retired, "the related costs and accumulated depreciation are removed from the accounts and any gain or loss is included in income."[5] Accumulated depreciation in 2017 was $179 thousand which brought property and equipment down from $319 thousand to $140 thousand. 2016's

accumulated depreciation of $116 thousand indicated a bigger drain of maintaining long-term assets.

Most of Finjan long term intangible assets come from the issues of patents. When issuing patents, Finjan "owns or possesses licenses to use its patents. The cost of maintaining the patents are expense as incurred." From February 2016, the Financial Accounting Standards Board issued Accounting Standards Update No. 2016-02 "Leases" "that requires a lessee to recognize the

4 Finjan, 2018 10-k, F-9
5 ibid

assets and liabilities that arise from operating leases….For leases with a term of 12 months or less, a lessee is permitted to make an accounting policy election by class of underlying asset not to recognize lease assets and lease liabilities" until December 15, 2018.[6] Their intangible assets amount to $7.748 million from 0 on December 31, 2017 out of $61 million in total assets.
The following long-term tangible assets decreased: Property and equipment (net) decreased from $203 thousand to $140 thousand, Investments from $2.745 million to $2.618 million and other non-current assets from $321 thousand to 0.[7] However, deferred income taxes, which are non-current assets, increased from 0 to $6.201 million. While every other long-term assets, that does not involve patents, decreased, Finjan relies on deferred income taxes due to its contractual obligations.

Despite Finjan's increased focus on intangible assets, its main key asset is from its cash and other cash equivalents which amount for $41.169 million out of $61.247 million in total assets.[8] Its policy "considers all highly liquid instruments with original maturities of three months or less when purchased to be cash equivalents. Included in cash and cash equivalents are demand deposits and money market accounts."[9] Although its cash assets surpass FDIC capital insurance policies, it has not experienced losses on its credit accounts.

Business deployment of Finjan's long-term assets generated revenues and produce products/services. With LTAs, grants from licenses to Finjan's patented cyber-security technology and settlements under the company's patent policies. To recognize the value of their revenue from its issue of patents, the arrangement with the licensee has been signed which is put into effect. For example, Finjan invested $5 million to Jerusalem Venture Partners (JVP), with

6 Finjan, 2018 10-k, F-13

-

-k, F

[7] Finjan 2018 10-k, F-15
[8] Finjan, 2018 10 k, F-34
[9] Finjan, 2018 10 -9

only $2.2 million remaining in its commitment to the company by December 31, 2017. In addition to its $2.8 million repaid from JVP, "On June 8, 2015, the Company received a cash distribution of $826 thousand as a portion of a gross entitlement of approximately $1.271 million from its investment in the JVP Fund.[10]" From that, the fair value of the patent would be recorded as revenue from their deployments of long-term assets.

While Finjan consolidated its revenue from payments of their licensees, the company must pay rents that they also utilized from outside companies. An example of its long-term debts comes from its rental of former headquarters in New York since 2013, paying in five-year annual payments of $139 thousand on a monthly basis of $12 thousand. Rent expense was $754 thousand in 2017 with an additional $36 thousand in rental payable, which is accounted in current liabilities of the balance sheet.[11] Finjan, with its successor Finjan mobile, also gained licensing agreements Avira, Inc. in which "Avira will provide its Virtual Private Network ("VPN") platform and technical support ("VPN Platform") to Finjan Mobile, and Finjan Mobile will use the VPN Platform as part of its Vital Security™ suite of product offerings….As of December 31, 2017, the Company has a $3.3 million contractual obligation due over the next 10 quarters."[12] Even though Finjan grants its own licensing patents to their licensees, it must also pay its own rents that they have which balances out.

-
-
-k, F

[10] Finjan, 2018 10 k, F-16
[11] Ibid [12] Ibid

**Three year time series analysis**

As revenues grew significantly from 2015 to 2017, gross margin % remained generally constant, (83% in 2015 and 2016, and 88% in 2017)[13]. Thus, gross margin $ naturally grew at a similar pace to revenue. Gross margin $ grew 1,048% from 2015-17, for a total of a $40.603 million increase in gross margin. The fact that as revenue grew, gross margin % remained generally constant means Finjan's business model and management can generate large revenue growth without large expenses, a good indicator that Finjan will be able to maintain profitability in the future. NOPAT grew during 2015-2017, from $-11.104 million to $11.526 million. Profit margin, adjusted profit margin, and operating profit margin also grew during these three years, from -269% to 45%, -237% to 23%, and -269% to 33% respectively. Thus, from 2015 to 2017, Finjan moved significantly in the right direction in every measure of profitability. However, this improvement is measured against a very poor base year of 2015. Thus, it is quite doubtful that Finjan can continue to grow at their current rates as they will likely soon run into economies of scale, slowing down their growth. In November of 2009 Finjan signed a confidential noncompete agreement with M86 Security; which expired in March of 2015. Finjan subsequently began to develop mobile security software and an advisory services section of its business[14]. These new developments contributed to the large losses and high expenses of 2015. 2015 was a poor year for Finjan, as they took a net loss of $12.602 million and all forms of profit margins were in the -200% range. Unusually high operating expenses contributed to Finjan's 2015 losses. 2015 operating expenses ($17.753 million) and revenues ($4.687 million) were above those in the previous and successive years[15]. Thus 2015 was an unusually unprofitably year. 2016 began

-

13 Finjan, 2018 10 k, 32-33
14 Finjan, 2016 10 k, 5
15 Finjan, 2015 10 -4

-
-
-k, F

the turnaround for Finjan after a poor 2015. Profit measures like operating profit margin, profit margin, and adjusted profit margin were a dismal 2% across the board. However, this represents a large increase from their poor 2015 performance mentioned above. 2016 $ value profitability measures tell a similar story to the % measures: small positive values, but a large improvement from 2015. While net income was small, its 2016 value is a massive step in the right direction after a $12.602 million net loss in 2015[16]. Finian's 2016 was a year of transition and should be remembered as a year where net income grew by nearly $13 million, not the year where net income was $350 thousand[17]. 2017 was by far the most profitable year for Finjan as gross margin $, net income, EBIT, and NOPAT all grew enormously[18]. Additionally, all forms of profit margin grew significantly into double digits from their 2% performance across the board in 2016. A significant portion (27%) of Finjan's $22.811 million net income in 2017 comes in the form of a $6.16 million income tax benefit due to $12.5 million in net operating losses carryforwards. Finjan believes these net operating losses carryforwards will provide tax benefits until 2026[19]. With more than a quarter of net income coming as repayment for previous losses, Finjan's 2017 profitability measures that involve net income (profit margin) should not be taken at face value. However, the other % and $ profitability measures remain impressive regardless.

From 2015 to 2017, cash flows from operations rose by an average of nearly $14 million per year, cash flows from investing declined by an average of $837 thousand per year, and cash flows from financing rose an average of $6.362 million per year. In total, Finjan saw a $23.664 million increase in cash and cash equivalents, less than the total cap raises during this period:

-
-
-

[16] Finjan, 2015 10-k, F-4
[17] Finjan, 2016 10 k, F-4
[18] Finjan, 2017 10 k, F-3
[19] Finjan, 2017 10 k, F-43

explained by a $11.404 million cash outflow in 2015 to expand their business[20]. During 2016 Finjan saw a net increase in cash and cash equivalents of $7.577 million, largely made up of a cash inflow of $6.808 million from financing activities[21]. On May 20, 2016 Finjan issued 102 thousand shares of Series A prefered stock to Halcyon LDRII in a non dilutive private placement resulting in a net cash infusion of $9.49 million[22]. Finjan also had a cash inflow of $1.328 million from operations during 2016, a $12 million increase from the previous year[23]. Thus, 2016 cash flows from operations tell the same story as 2016 profit measures: modest figures showing progress. In 2017 Finjan saw a large increase in net cash provided by operating activities; driven by $22.811 million in net income. This large cash inflow is a continuation of previous increases that should reassure investors that the business model works, and that the management team has implemented it effectively. The $2 million cash outflow from investing activities in 2017 is the first of four payments totaling $8.5 million to IBM in exchange for a series of patents[24]. During 2017 Finjan redeemed the remainder of their Series A Preferred shares issued the year prior for a cash outflow of $13.778 million. However, Finjan once again raised capital through stock issuances, generating $14.375 million and $11.952 million from issuances of Series A-1 Preferred shares and common shares respectively[25]. Finjan cannot continue to issue stock indefinitely, and some will dismiss Finjan's growth as due to a massive and unsustainable infusion of capital totalling $35.817 million. However, the combination of strong growth in net cash provided by operations and the expiration of Finjan's non-compete agreement in 2015 indicate that the capital infusions did not drive the impressive returns, but rather that investors

-
-
-k, F

[20] Finjan, 2015 10-k, F-6
[21] Finjan, 2016 10-k, F-6
[22] Finjan, 2016 10-k, 36
[23] Finjan, 2016 10 k, F-6
[24] Finjan, 2017 10 k, 10
[25] Finjan, 2017 10 -6

flocked to the returns. Furthermore, the swift and total redemption of all Series A Preferred shares means Finjan essentially took out a one year loan before quickly paying it back, indicating strong financial health of the company.

Unit and price information is not applicable to Finjan's patent troll business model. Finjan saw massive increases in revenues from 2015-2017 of $45.797 million total. Each year their revenue more than doubled, increasing 977% over the three year period. From 2015-2017 more than half Finjan's revenue came from lawsuits or settlements of lawsuits to enforce their patents. This composition of revenue indicates Finjan's reliance on current patent law and the quality of their legal council. In 2015 Finjan won a $39.5 million lawsuit but have not (as of the 2017 10-k) recognized or received any of this revenue, leaving doubts about its collectability[26]. During 2016 total operating expenses fell, however, Finjan believes this reduction in SGA expenses came from the timing of their lawsuits rather than their ability to reduce costs on their own[27]. Importantly, as revenue grew 175% in 2017, cost of revenues only grew 97%; meaning Finjan has yet to hit economies of scale, and will likely continue to see strong growth in profitability measures during the coming years before slowing down. However, the inability of Finjan to control the timing and amount of their expenses should worry investors.

The slight increase in LTA from 2015 to 2016 stemmed from funding of Finjan's $5 million commitment to Jerusalem Venture Partners (JVP)[28]. The large increase in LTA from 2016 to 2017 was due to Finjan realizing $6.201 million in a deferred income tax asset as well as the

-
-
-k, F

$8.5 million acquisition of patents from IBM. In 2015 and 2016, the vast majority of LTA were investments in a venture capital fund run by JVP. During 2015 JVP returned $1.271 million

26 Finjan, 2016 10 k, 7
27 Finjan, 2016 10 k, 39
28 Finjan, 2016 10 -13

-
-
-k, F

to Finjan. This return came when Finjan had only invested $1 million, yielding a 127% return over two years. This impressive return demonstrates JVP is a valuable long term asset that will likely provide cash to Finjan in the future. Importantly, 11 of the 27 patents held by the company in 2015 expired by the end of 2017, yet Finjan's revenue and profitability grew over this period, and so did the number of patent owned by Finjan. These facts indicate Finjan developed or acquired profitable patents over this three year period that can continue to provide revenue in future years. During 2015 and 2016 long term liabilities were less than 5% and 3% of total liabilities, respectively. Neither the 2016 nor 2015 10-K elaborate on the makeup of these more or less insignificant liabilities. In 2017 long term liabilities rose to $5.5 million, comprising almost 40% of total liabilities. This dramatic rise in long term liabilities is due to the long term portion of Finian's commitment to pay $8.5 million over 4 years to IBM in exchange for a series of patents.

A $17.606 million accumulated deficit provides the most significant drag on Equity in 2015. Total and average (of 2015-2014) equity were $6.342 and $12.233 million respectively[29]. This large disparity between actual and average equity in 2015 is due to high equity 2014. When the inflated average equity in 2015 is compared to the average equity of 2016, there appears to be a $8.619 million drop. However, the true drop in equity is $5.456 million, almost exclusively driven by a reduction in APIC due to $6.789 million accretion of their newly issued Series A Preferred stock[30]. This accretion is an increase in the book value of Series A Preferred stock and represents a part of the costs of the capital infusion from Series A Preferred. In this way, the accretion is similar to interest payments on debt, however, there is no actual outflow of cash.

-

-k, F

29 Finjan, 2016 10 k, F-3
30 Finjan, 2016 10 -3

Finjan pays no dividends from 2015-2017 and thus net income is equal to retained earnings[31]. So while 2016 net income of $350 thousand was a large increase from net income in 2015, it did very little to reduce the $17.606 million accumulated deficit that Finjan began 2016 with. This large accumulated deficit is a significant drag on 2016 equity. In 2017, equity increased for the first time in this three year time series, rising by $27.64 million. This large increase in equity was partially driven by 2017 net income of $22.811 million, allowing Finjan to eliminate their previous years accumulated deficit. Sale of common stock also provided the company with nearly $12 million, further contributing to Finian's large growth in equity[32]. The composition of Finjan's equity tells a good and similar story to other ratios and metrics explored above. Each year the composition of Finjan's equity moved in the right direction; accumulated deficits decreased and were eventually eliminated, meanwhile, the company raised more and more money through stock issuances, providing a helpful cash infusion.

Despite Finjan's poor performance in 2015, liquidity measures indicated no liquidity trouble for the company. Finjan's current and quick ratio were 2.35 and 2.23 respectively. The small difference between current and quick ratios (A/R=0) represents a cash heavy composition of current assets that is good for liquidity. Furthermore, Finjan's long term liabilities were minimal, and current assets more than doubled total liabilities, another indicator of a liquid business. As revenue and profits rose in 2016, so too did liquidity. Finjan's current ratio rose to 4.95 and their quick ratio increased to 3.75. Current assets almost tripled while total liabilities only rose by 37%. Finjan's liquidity measures reaffirm the strong growth made by the company from 2015-2017. In 2017 Finjan's liabilities grew by almost $10 million, rising 249%[33].

-

31 Finjan, 2017 10 k, 25
32 Finjan, 2017 10 k, F-5
33 Finjan, 2017 10 -6

However, this is no cause for concern, since current and non-current assets grew as well. The current and quick ratios rose to 5.39 and 5.3 respectively. Finjan's ability to increase liquidity while also increasing revenue and net income stems from their large capital raises. These large capital raises allow Finjan to finance growth without taking on debt. From 2015-2017 the quick and current ratios rose each year; meaning that future debt payments will not bring Finjan's growth to a screeching halt.

From 2015-2017 the company's financial health strengthened. All measures of profitability and revenue grew significantly each year. The company's net cash provided by operating activities and total assets also saw impressive yearly growth. Importantly, none of this growth was financed by debt; and thus liquidity and equity grew over the three year period as well. Finjan states that their revenue increases were driven primarily by increased courtroom success in enforcing infringement on their patents[34]. Finjan's courtroom based business model will likely continue to yield success for the company. From 2015-2017 the number of lawsuits Finjan brought and the revenues from those lawsuits increased, this trend will undoubtedly attract higher quality attorneys to represent Finjan; thus increasing their potential for revenue. Additionally, the number of tech companies is on the rise, and this trend will continue to be for the foreseeable future. While this can be seen as a source of competition for Finjan, it can also be seen as a source of revenue; as more and more companies can possibly infringe on Finjan's patents.

-
-
-k, F

## Cross-Sectional Analysis

Like Finjan, NTIP also has litigations pending against companies including Microsoft as a result of not receiving royalty payments[35]. Based on CF Ops and Free Cash Flows, FNJN is in

[34] Finjan, 2016 10-k, 39
[35] NTIP, 2017 10 K, 11

-

a growth phase. NTIP is in a steady growth phase, IDCC and DLB are in Maturity. The CF Ops metric and FCF metric agrees with the founding dates of these companies. DLB is more stable in its CF Ops than IDCC. FNJN has had the most risky and least consistent profit margins ranging from -269% to 45% presumably as a result of its age and growth position.

IDCC has the risks of negative free cash flows[36], inability to pay of debts[37], liable to changes in dividends paid which would drive investors away[38]. Luckily, free cash flows have been positive since 2015. The risk of negative cash flows comes from its heavy spending business model: the company needs cash to finance its infrastructure[39]. Its product depends on people's belief in the safety of mobile devices with its RF frequency risks[40]. Problematically, 18% of its "recurring revenue" comes from Huawei and it is in lawsuit by "Chinese Anti Monopoly" law from "making proposals for royalties from Huawei that the court believed were excessive[41]".

As a result of its losing streak at lawsuits, FNJN had negative economic profit and profit margin until 2016. FNJN has had a cyclical pattern of revenue as a result of its businesses model of patent trolling. Every company peaked in 2016 but FNJN improved the most in 2016 as a result of their motto.

According to FNJN's 10-K its research and development expenses are fluctuate and also continue to vacuum its revenue. IDCC and the other competitors of FNJN also have risks that affect their profit margins, but those listed for example on IDCC's 10-K including "currency

36 IDCC, 2017 10-K, 13
37 IDCC, 2017 10-K, 21
38 ibid
39 IDCC, 2017 10 K, 14
40 IDCC, 2017 10-K, 5

-

-

41 IDCC, 2017 10 K, 24

fluctuations[42]" given its overseas business, litigation risk, quality of services, and liabilities on products it sells with warranties have not mattered[43].

IDCC has a current ratio of around 3 making it prepared to pay liabilities and undergo expected regular litigations. FNJN is a "patent troll" as a result of its inability to receive patent royalties, in the words of Linwood Downs. Unfortunately, FNJN counts litigation expenses resulting from patent infringements as its SGA expense, which makes up more than half to four times its revenue over the past three years. FNJN's competitor, IDCC, has an advantage over FNJN, in that it does not need to spend as much money on litigation as FNJN. The users of IDCC's patent pay their royalties that account for one fifth of its revenue. Whereas FNJN has had ongoing expense to litigate to protect its patents, IDCC has had a decrease in operating costs resulting from litigation of $"16.1" million. In perspective, FNJN has had negative revenue resulting from litigation operating expenses whereas IDCC's revenue has never been affected severely by operating expenses that continue to decrease by about one million from 2015 to 2017[44]. Because younger FNJN has less cash, its failure to protect its patents may be more significant than that of IDCC. Although FNJN is less risky than IDCC according to its beta value, IDCC is less risky than FNJN as a result of its longer history and more stable profit margins and its non zero economic profit. Although IDCC has many current advantages over FNJN in its stable profitability, this advantage is short-lived. IDCC's 2016 10-K noted that FASB mandated it to estimate "royalty revenues[45]" and change their timing of recording. This mandate will devalue IDCC's source of revenue and make its periodic income seem artificially

42 IDCC, 2017 10-k, 19
43 Ibid

44 IDCC, 2017 10-k 56
45 IDCC, 2016 10-K, 12

unstable[46]. IDCC lists in its 2016 report that it needs to lure qualified engineers and technology people to effectively run its business. The long-term costs of re-training, training new employees, offering competitive employment packages in their opinion pales in comparison to the profitability these bring. If many companies such as IDCC exists, then this will bring good incentive in the education industry as engineering and technology education will be more useful. IDCC with enough capital may consider marketing outreach to universities, high schools to increase education in its fields in Science Technology Engineering Mathematics as well as fund local education in its district. If a shortage of women exist, it may run valuable marketing campaigns targeting educating women in a traditionally male dominated fields. These corporate data suggest a trend towards social equality or competition within companies may result in increased social equality and opportunity if seen in an optimistic light. Companies have the power to change culture. The company with the highest current ratio that also requires educated personnel has the most capacity to hire bright minds. NTIP would have more competitive advantage than all with its highest current ratio with excess cash to hire the most qualified people however the effect of hiring talented minds takes time to develop. In the meantime, Dolby seems the best option to invest but NTIP seems promising.

Out of the companies, Dolby is the company with the best overall financial and operation health. According to the table that accounted its financial statements for the past three years, Dolby has stable profit margins that generates a revenue in the billions, the only company in ten figures.

[46] Ibid

| Liquidity Table of FNJN (Thousands) | Dec. 31, 2017 | Dec. 31, 2016 | Dec. 31, 2015 |
|---|---|---|---|
| Revenue$ | 50,484 | 18,387 | 4,687 |
| Revenue Growth% | 175% | 292% | |
| Gross Margin | 44,476 | 15,350 | 3,873 |
| Gross Margin % | 88% | 83% | 83% |
| EBIT | 16,65[1] | 353 | -12,597 |
| Operating Profit Margin% | 33% | 2% | -269% |
| Net Income | 22,811 | 350 | -12,602 |
| Profit Margin% | 45% | 2% | -269% |
| Risk Free Rate%[1] | 2.46% | 2.48% | 2.27% |
| Beta[2] | -0.47 | -.47 | -.47 |
| Market Premium% | 5% | 5% | 5% |
| Re%: Equity Cost of Capital% | 0.1% | 0.1% | -0.1% |
| Re$: $Equity Cost of Capital | 16.2 | 4.7 | -9.8 |
| $Economic Profit (equity) | 22,795 | 345 | -12,592 |
| %Economic Profit (equity) | 45.2% | 1.9% | -268.7% |
| Average Equity | 14,706 | 3,614 | 12,233 |
| Tax Rate% | 0% | 0.85% | 0.04% |
| NOPAT | 16,651 | 350 | -12,592 |
| Average Assets | 39,839 | 13,753 | 19,950 |
| Adjusted Profit Margin% | 33% | 2% | -269% |
| Asset Turnover Ratio | 1.27 | 1.34 | .23 |
| ROA% | 41.8% | 2.5% | -63.1% |
| Capital Structure Leverage Ratio | 2.71 | 3.81 | 1.63 |
| ROE% | 155.1% | 9.7% | -103% |
| Invested Capital | 14,706 | 3,614 | 12,233 |
| ROIC% | 113.2% | 9.7% | -102.9% |
| Depreciation Expense | 63 | 63 | 50 |
| Accumulated Depreciation | 179 | 116 | 68 |
| Cash ROIC% | 112.3% | 11.1% | -102% |
| LTD | 0 | 0 | 0 |
| $\frac{\textbf{LTD}}{\textbf{LTD} + \textbf{Equity}}\%$ | 0% | 0% | 0% |
| Interest Expense | 0 | 0 | 0 |
| After Tax Interest Rate | 0 | 0 | 0 |
| $\frac{\textbf{Equity}}{\quad}\%$ | 100% | 100% | 100% |

[1] Wall Street Journal

[2] Finviz

| | | | |
|---|---|---|---|
| **($WAAC)(IC)WACC%** | 0.01% | 0.01% | -0.01% |
| **\$Economic Profit$_{Firm}$** | 16,635 | 345 | -12,582 |
| **%Economic Profit$_{Firm}$** | 113.1% | 9.6% | -102.9% |
| Current Ratio | 5.3948 | 4.9525 | 2.3544 |
| Tie Ratio: Interest Coverage | N/A | | |
| CF Ops | 16,586 | 1,328 | -11,259 |
| LT Liabilities | 5,500 | 119 | 130 |
| **CF OPs** | 3.02 | 11.16 | -86.61 |
| **LTD + Equity** | | | |
| MV/BV[1] | 4.3 | | 3.57 |
| WACC% | 0.1% | 0.1% | -0.1% |

**LT Liabilities**

| Liquidity Table of DLB (Millions) | Dec. 31, 2017 | Dec. 31, 2016 | Dec. 31, 2015 |
|---|---|---|---|
| Revenue$ | 1,080 | 1,030 | 970.64 |
| Revenue Growth% | 4.85% | 6.12% | |
| Gross Margin$ | 963.15 | 916.76 | 875.82 |
| Gross Margin % | 89% | 89% | 90% |
| EBIT | 261 | 233 | 214 |
| Operating Profit Margin% | 24.17% | 22.62% | 22.05% |
| Net Income | 201.8 | 185.86 | 181.39 |
| Profit Margin% | 18.69% | 18.04% | 18.69% |
| Risk Free Rate% | 2.46% | 2.48% | 2.27% |
| Beta | 0.74 | 0.74 | 0.74 |
| Market Premium | 5% | 5% | 5$ |
| Re%: Equity Cost of Capital% | 6.16% | 6.18% | 5.97% |
| Re$: $Equity Cost of Capital | 405.09 | 270.88 | 198.05 |
| $Economic Profit (equity) | -203.29 | -85.02 | -16.66 |
| %Economic Profit (equity) | 3.63% | 3.61% | 4.25% |
| Tax Rate | 23 | 20 | 15 |
| NOPAT | 201.8 | 185.8 | 181.3 |
| Average Assets | 2,429 | 3,383 | 2,058 |
| Adjusted Profit Margin | 19% | 18% | 19% |
| Asset Turnover Ratio | 0.44 | 0.30 | 0.47 |
| ROA% | 8.31% | 5.49% | 8.81% |
| Average Equity | 2,061 | 1,897 | 1,774 |
| Capital Structure Leverage Ratio | 1.18 | 1.78 | 1.16 |
| ROE% | 9.79% | 9.79% | 10.22% |

[1] Zacks.com

| | | | |
|---|---|---|---|
| Invested Capital | 2,062 | 1,898 | 1,774 |
| ROIC% | 9.79% | 9.79% | 10.22% |
| Depreciation Expense | 53 | 52 | 48 |
| Accumulated Depreciation | 305 | 251 | 202 |
| Cash ROIC% | 11% | 11% | 12% |
| LTD | 0 | 0 | 0 |
| $\frac{\textbf{LTD}}{\textbf{LTD + Equity}}$ | 0 | 0 | 0 |
| Interest Expense | 0.13 | 0.13 | 0.18 |
| After Tax Interest Rate | 0 | 0 | 0 |
| $\frac{\textbf{Equity}}{\textbf{LTD + Equity}}\%$ | 100% | 100% | 100% |
| MV/BV | 3.19 | 2.31 | 1.87 |
| WACC% | 6.16% | 6.18% | 5.97% |
| **($WAAC)(IC)WACC%** | 127 | 117 | 106 |
| $\textbf{\$Economic Profit}_{\textbf{Firm}}$ | 74.81 | 68.59 | 75.48 |
| $\textbf{\%Economic Profit}_{\textbf{Firm}}$ | 3.63% | 3.61% | 4.25% |
| Current Ratio | 4.12 | 3.57 | 4.38 |
| Tie Ratio: Interest Coverage | 2,007.7 | 1,792.3 | 1,188.9 |
| CF Ops | 371.05 | 356.84 | 309.38 |
| LT Liabilities | 151 | 125 | 107 |
| $\frac{\textbf{CF OPs}}{\textbf{LT Liabilities}}$ | 2.46 | 2.85 | 2.89 |

| Liquidity Table of IDCC (Millions) | Dec. 31, 2017 | Dec. 31, 2016 | Dec. 31, 2015 |
|---|---|---|---|
| Revenue$ | 532.94 | 665.85 | 441.44 |
| Revenue Growth% | -19.96% | 50.84% | |
| Gross Margin | 421 | 552 | 321 |
| Gross Margin % | 88% | 83% | 73% |
| EBIT | 292 | 422 | 181 |
| Operating Profit Margin% | 55% | 63% | 41% |
| Net Income | 174 | 309 | 119 |
| Profit Margin% | 32% | 46% | 27% |
| Risk Free Rate% | 2.46% | 2.48% | 2.27% |
| Beta | 1.07 | 1.07 | 1.07 |
| Market Premium% | 5% | 5% | 5% |
| Re%: Equity Cost of Capital% | 7.81% | 7.83% | 7.62% |
| Re$: $Equity Cost of Capital | 211 | 247 | 134 |
| Invested Capital | 1,099 | 910 | 758 |
| ROIC% | 15.85% | 33.95% | 15.7% |
| Depreciation Expense | 4 | 4.1 | 3.8 |
| Accumulated Depreciation | 28 | 24 | 51 |
| Cash ROIC% | 16% | 34% | 15% |
| LTD | 285 | 272 | 260 |
| $\frac{\textbf{LTD}}{\textbf{LTD + Equity}}$ | 0.1 | 0.08 | 0.13 |
| Interest Expense | 17 | 21 | 30 |
| After Tax Interest Rate% | 3.7% | 5.7% | 7.7% |
| $\frac{\textbf{Equity}}{\textbf{LTD + Equity}}$ | 0.90 | 0.92 | 0.87 |
| MV/BV | 3.33 | 4.96 | 3.53 |
| WACC% | 7.42% | 7.66% | 7.63% |
| **(\$WAAC)(IC)WACC%** | 82 | 70 | 58 |
| $\$\textbf{Economic Profit}_{\textbf{Firm}}$ | 96 | 239 | 61 |
| $\%\textbf{Economic Profit}_{\textbf{Firm}}$ | 8.43% | 26.29% | 8.09% |
| Current Ratio | 3.7 | 2.87 | 2.53 |
| Tie Ratio: Interest Coverage | 16 | 19 | 5 |
| CF Ops | 316 | 434 | 116 |
| LT Liabilities | 605 | 548 | 553 |
| **CF OPs** | 0.52 | 0.79 | 0.21 |
| \$Economic Profit (equity) | -37 | 61 | -14 |
| %Economic Profit (equity) | 13% | 65% | 6% |
| Average Equity | 814 | 638 | 499 |
| Tax Rate% | 40% | 27% | 34% |
| NOPAT | 174 | 309 | 119 |
| Average Assets | 1,791 | 1,601 | 1,333 |
| Adjusted Profit Margin% | 33% | 46% | 27% |
| Asset Turnover Ratio | 0.30 | 0.42 | 0.331 |
| ROA% | 9.73% | 19.30% | 8.94% |
| Capital Structure Leverage Ratio | 2.20 | 3.81 | 1.63 |
| ROE% | 21.41% | 73.44% | 14.58% |

**LT Liabilities**

| **Liquidity Table of NTIP (Thousands)** | **Dec. 31, 2017** | **Dec. 31, 2016** | **Dec. 31, 2015** |
|---|---|---|---|
| Revenue$ | 16,451 | 65,088 | 16,565 |
| Revenue Growth% | -74.72% | 292.92% | |
| Gross Margin | 11,481 | 39,294 | 11,059 |
| Gross Margin % | 30% | 40% | 33% |
| EBIT | 6,229 | 32,161 | 3,985 |
| Operating Profit Margin% | 38% | 49% | 24% |
| Net Income | 4,122 | 23,227 | 4,107 |
| Profit Margin% | 25% | 36% | 25% |
| Risk Free Rate% | 2.46% | 2.48% | 2.27% |
| Beta | -.35 | -.35 | -.35 |
| Market Premium% | 5% | 5% | 5% |
| Re%: Equity Cost of Capital% | 0.71% | 0.73% | 0.52% |
| Re$: $Equity Cost of Capital | .421 | .598 | .276 |
| $Economic Profit (equity) | 3,700 | 22,628 | 3,831 |
| %Economic Profit (equity) | 6.8% | 43.2% | 13.8% |
| Average Equity | 54,412 | 52,872 | 28,690 |
| Tax Rate% | 34% | 28% | 0% |

| NOPAT | 4,122 | 23,227 | 3,985 |
|---|---|---|---|
| Average Assets | 57,294 | 57,597 | 30,381 |
| Adjusted Profit Margin% | 25% | 36% | 24% |
| Asset Turnover Ratio | 0.29 | 1.13 | 0.55 |
| ROA% | 7.19% | 40.33% | 13.12% |
| Capital Structure Leverage Ratio | 1.05 | 1.09 | 1.06 |
| ROE% | 7.58% | 43.93% | 14.32% |
| Invested Capital | 54,412 | 52,872 | 28,690 |
| ROIC% | 7.6% | 43.9% | 13.9% |
| Depreciation Expense | 0 | 0 | 0 |
| Accumulated Depreciation | 0 | 0 | 0 |
| Cash ROIC% | 8% | 44% | 14% |
| LTD | 0 | 0 | 0 |
| $\frac{\textbf{LTD}}{\textbf{LTD + Equity}}$ | 0 | 0.0 | 0.0 |
| Interest Expense | 0% | 0% | 0% |
| After Tax Interest Rate | 0% | 0% | 0% |
| $\frac{\textbf{Equity}}{\textbf{LTD + Equity}}\%$ | 100% | 100% | 100% |
| MV/BV | 1.09 | 1.55 | 1.85 |
| WACC% | 0.71% | 0.73% | 0.52% |
| $(\$\textbf{WAAC})(\textbf{IC})\textbf{WACC}\%$ | .386 | .386 | .149 |
| $\$\textbf{Economic Profit}_{\textbf{Firm}}$ | 3,735 | 22,841 | 3,835 |
| $\%\textbf{Economic Profit}_{\textbf{Firm}}$ | 6.8% | 43.2% | 13.3% |
| Current Ratio | 23.31 | 11.81 | 13.83 |
| Tie Ratio: Interest Coverage | N/A | N/A | N/A |
| CF Ops | 6,774 | 29,906 | 5,633 |
| LT Liabilities | 0 | 0 | 0 |
| $\frac{\textbf{CF OPs}}{\textbf{LT Liabilities}}$ | N/A | | |

Documents